\documentclass[11pt, letterpaper]{template}
\usepackage{graphicx} 
\usepackage[utf8]{inputenc} 
\usepackage[T1]{fontenc}    
\usepackage{url}            
\usepackage{booktabs}       
\usepackage{amsfonts}       
\usepackage{nicefrac}       
\usepackage{microtype}      
\usepackage{xcolor}         

\usepackage{hyperref}
\hypersetup{
    breaklinks,
    colorlinks,
    citecolor=blue
}

\usepackage{microtype}
\usepackage{graphicx}
\usepackage{tabularx}
\usepackage{booktabs} 
\usepackage[utf8]{inputenc} 
\usepackage{url}
\usepackage{booktabs}    
\usepackage{amsfonts}     
\usepackage{nicefrac}       
\usepackage{microtype}     
\usepackage{utfsym}
\usepackage{float}
\usepackage{color}
\usepackage[most]{tcolorbox}
\usepackage{amsmath}
\usepackage{amsthm}
\usepackage{multirow}
\usepackage{multicol}
\usepackage{amssymb}
\usepackage{longtable}
\usepackage{soul}
\usepackage{subcaption}
\usepackage{bbding}
\usepackage{indentfirst}
\usepackage{cleveref}
\usepackage{pifont}
\usepackage{diagbox}
\usepackage{graphicx}
\usepackage{amsmath,amssymb,amsfonts}
\usepackage{textcomp}
\usepackage{xcolor}
\usepackage{makecell} 

\usepackage[authoryear, round]{natbib}
\newcommand{\myname}{TrinityGuard}
\renewcommand{\cite}{\citep}
\newcommand\blfootnote[1]{%
  \begingroup
  \renewcommand\thefootnote{}\footnote{#1}%
  \addtocounter{footnote}{-1}%
  \endgroup
}
\title{\myname: A Unified Framework for Safeguarding Multi-Agent Systems}

\author[]{Kai Wang*, Biaojie Zeng*, Zeming Wei*, Chang Jin, Hefeng Zhou, Xiangtian Li, Chao Yang, Jingjing Qu, Xingcheng Xu, Xia Hu \\
Shanghai AI Laboratory\\
\url{https://github.com/AI45Lab/TrinityGuard}}

\date{February 2026}

\begin{document}

\begin{abstract}
With the rapid development of LLM-based multi-agent systems (MAS), their significant safety and security concerns have emerged, which introduce novel risks going beyond single agents or LLMs. Despite attempts to address these issues, the existing literature lacks a cohesive safeguarding system specialized for MAS risks. In this work, we introduce \textbf{TrinityGuard}, a comprehensive safety evaluation and monitoring framework for LLM-based MAS, grounded in the OWASP standards. Specifically, TrinityGuard encompasses a three-tier fine-grained risk taxonomy that identifies 20 risk types, covering single-agent vulnerabilities, inter-agent communication threats, and system-level emergent hazards. Designed for scalability across various MAS structures and platforms, TrinityGuard is organized in a trinity manner, involving an MAS abstraction layer that can be adapted to any MAS structures, an evaluation layer containing risk-specific test modules, alongside runtime monitor agents coordinated by a unified LLM Judge Factory. During Evaluation, TrinityGuard executes curated attack probes to generate detailed vulnerability reports for each risk type, where monitor agents analyze structured execution traces and issue real-time alerts, enabling both pre-development evaluation and runtime monitoring. We further formalize these safety metrics and present detailed case studies across various representative MAS examples, showcasing the versatility and reliability of TrinityGuard. Overall, TrinityGuard acts as a comprehensive framework for evaluating and monitoring various risks in MAS, paving the way for further research into their safety and security.
\end{abstract}

\blfootnote{$^*$ Equal Contribution}

\maketitle

\section{Introduction}
\label{sec:intro}

Large Language Model (LLM)-based agents have evolved from single-turn assistants into autonomous entities that plan, invoke tools, and reason over extended horizons~\cite{zheng2025deepresearcher,huang2025deep,gao2025efficient}. Recently, composing several such agents into \emph{multi-agent systems} (MAS) enables collaborative workflows in software engineering~\cite{he2025llm}, scientific discovery~\cite{ghafarollahi2025sciagents}, financial analysis~\cite{du2025retrieval} with orchestration frameworks, such as AutoGen~(AG2)~\cite{wu2024autogen}, making deployment increasingly accessible. As the use of these systems becomes more widespread, concerns about their safety and security have also emerged~\cite{deng2025ai,ma2026safety,wei2025position}. Beyond the known risks in individual LLMs or agents, such as producing harmful~\cite{zou2023universal,wei2026jailbreak} or hallucinated~\cite{ji2023towards,bang2025hallulens} contents, MAS pose significantly more novel risks to be addressed.

These risks associated with MAS are inherently \emph{hierarchical}. We propose a three-tier taxonomy of MAS risks as follows. First, at the \textbf{individual-agent} level, each agent inherits a conventional attack surface like prompt injection~\cite{liu2023prompt} and jailbreaking~\cite{zou2023universal} as catalogued by the OWASP Top~10 for LLM Applications~\cite{owasp_llm_top10}. Second, at the \textbf{inter-agent communication} level~\cite{yan2025beyond}, multi-agent interaction introduces qualitatively new risks: malicious instructions can {propagate} through message channels~\cite{gu2024agent}, factual errors can be {amplified} by group dynamics~\cite{wu2025monitoring}, and agents can {spoof} one another's identities to escalate privileges~\cite{de2025open}. Finally, at the \textbf{system} level, the collective behavior of the MAS may exhibit \emph{emergent} phenomena like agent collusion~\cite{ghaemi2025survey} dynamics, where no individual agent would produce in isolation.
While these new risks have attracted more attention from the research community, e.g., the OWASP Agentic AI Top 10 standard has begun to codify these threats~\cite{owasp_agentic_2026}, standardized tooling for systematically evaluating and monitoring them across heterogeneous MAS deployments remains largely unaddressed.

To address the challenges above, we present \textbf{TrinityGuard}, a comprehensive and scalable framework for safeguarding MAS risks. First, to make TrinityGuard scalable to various platform-agnostic MAS when evaluating these risks, we build a three-layer decoupling safeguarding architecture. First, the MAS framework layer abstracts over heterogeneous orchestration frameworks through a unified \texttt{BaseMAS} abstraction interface (e.g., AG2/AutoGen, LangGraph, CrewAI, and others), fully decoupling safety logic from any specific MAS library. Then, the MAS Intermediary Layer provides framework-agnostic primitives for both test-time intervention and runtime interception. Finally, at the top layer, various risk-specific test modules paired with corresponding runtime monitor agents are deployed for judging these risks.

We then hierarchically study and gather 20 popular MAS risk types into three tiers that mirror the layered nature of MAS threats:
\begin{enumerate}
\item \textbf{Risk Tier~1 (RT1):} 8 single-agent atomic risks (e.g., prompt injection, jailbreaking, hallucination).
\item \textbf{Risk Tier~2 (RT2):} 6 inter-agent communication risks (e.g., malicious propagation, misinformation amplification, identity spoofing).
\item \textbf{Risk Tier~3 (RT3):} 6 system-level emergent risks (e.g., group hallucination, rogue agent).
\end{enumerate}

During evaluation, TrinityGuard~generates and executes attack test cases combining curated static payloads with LLM-synthesized adaptive probes against each agent (Tier~1), communication channel (Tier~2), and full execution trajectory (Tier~3), producing structured vulnerability reports with per-entity, per-risk pass rates and severity profiles, and gathering them as a comprehensive evaluation report. We also present evaluations and case studies on a group of MAS to validate TrinityGuard. Our principal contributions are:
\begin{enumerate}
\item An \textbf{open-source, extensible framework TrinityGuard} whose three-layer architecture cleanly decouples MAS abstraction, evaluation scaffolding, and risk-specific safety logic, enabling plug-in extension of risk modules and MAS platform support with minimal integration effort.
\item A \textbf{three-tier risk taxonomy} that organizes 20 MAS risk types across single-agent, inter-agent, and system-level dimensions, grounded in the OWASP standards and formalized with quantitative per-entity, per-risk safety metrics.
\item A \textbf{unified evaluation pipeline} that combines curated static payloads with LLM-synthesized adaptive probes targeting each risk type and tier, and produces structured vulnerability reports with pass rates and severity profiles, as demonstrated by comprehensive evaluation and case studies.
\end{enumerate}

The remainder of the paper is organized as follows.
Section~\ref{sec:method} presents the three-layer evaluation methodology.
Section~\ref{sec:metrics} formalizes the security metrics and the three-tier risk taxonomy.
Section~\ref{sec:eval} presents experimental evaluation.
Section~\ref{sec:discuss} discusses related work, limitations, and future directions.
Section~\ref{sec:conclusion} concludes.

\section{Evaluation Methodology}
\label{sec:method}

\begin{figure}
    \centering
    \includegraphics[width=0.9\linewidth]{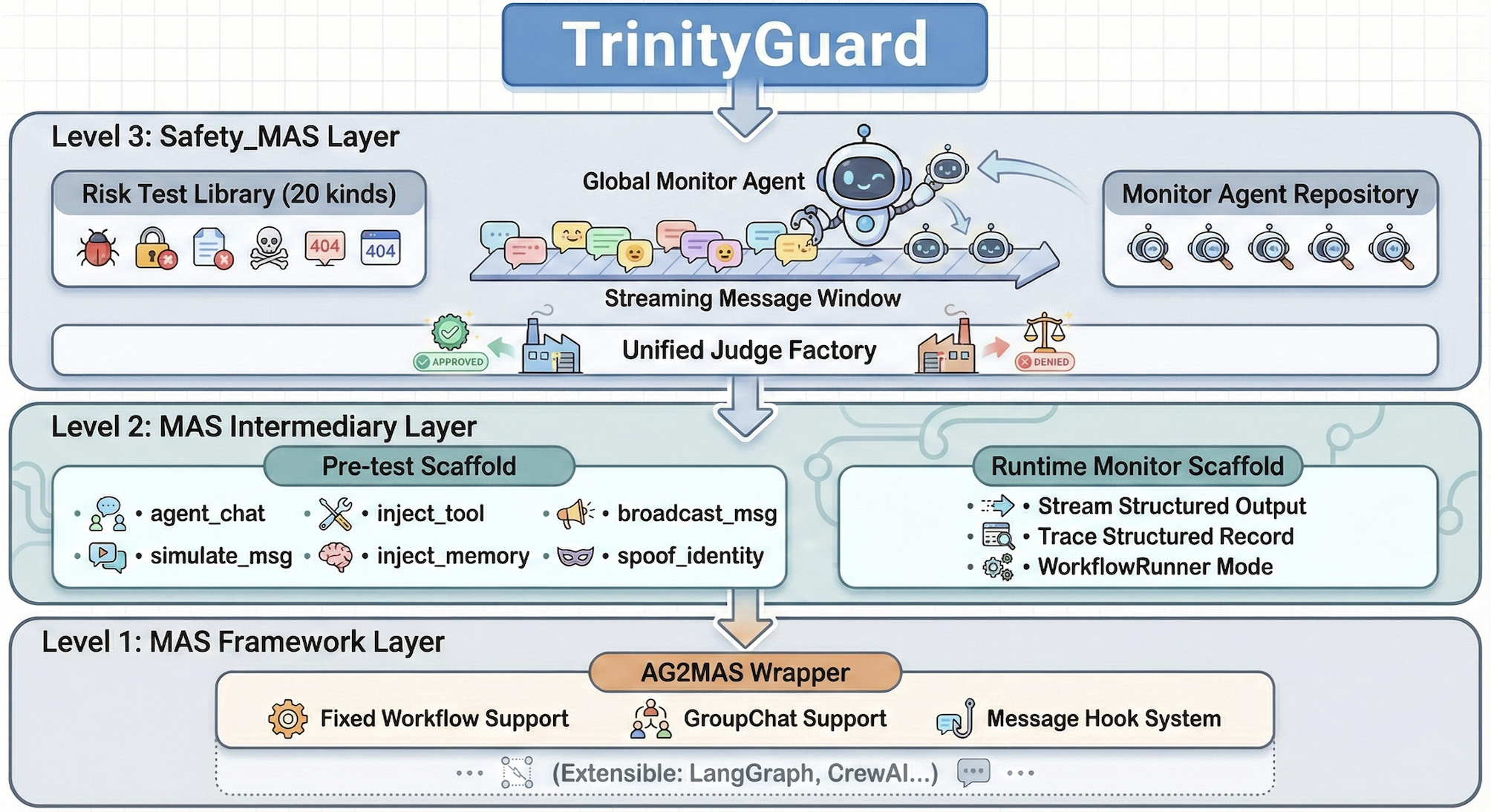}
    \caption{The overall architecture design of TrinityGuard.\vspace{-10pt}}
    \label{fig:level}
\end{figure}

This section presents the high-level design of TrinityGuard's evaluation methodology.
To systematically assess the risks across heterogeneous MAS deployments, we adopt a three-layer architecture that cleanly separates platform-specific concerns from safety evaluation logic.
Figure~\ref{fig:level} illustrates the overall architecture.

\subsection{Level~1: MAS Abstraction Layer}
\label{sec:method:l1}

As stated, our first design principle is that all safety evaluation logic should be agnostic to the specific MAS orchestration framework in use (e.g., AG2 and others). Level~1 achieves this by defining a unified abstraction over heterogeneous MAS platforms through a minimal interface: enumerating agents, routing messages, and executing tasks. Crucially, this abstraction also standardizes a message-hook mechanism that allows upper layers to intercept and inspect messages in transit, which is a prerequisite for both test-time intervention (Level~2) and runtime monitoring (Level~3). Based on this layer, adapting to a new MAS framework requires only implementing this abstraction layer, while all safety evaluation functionality is inherited without modification.

\subsection{Level~2: Intermediary Layer}
\label{sec:method:l2}

Level 2 bridges the platform abstraction and the safety logic by providing two families of framework-agnostic primitives: test-time intervention and runtime observation. These primitives serve as a transitional platform for connecting the safety evaluation modules and the abstracted MAS structure.

\paragraph{Test-time intervention primitives.}
Evaluating MAS security requires the ability to inject controlled adversarial stimuli into a running system. Level~2 provides a suite of \textit{intervention primitives} that cover the principal attack surfaces, like direct messaging to a target agent, fabricating inter-agent messages, poisoning agent memory stores, registering malicious tools, and impersonating trusted agents. These primitives are controlled by Level~3 risk tests to realize the attack scenarios defined in Section~\ref{sec:metrics}.

\paragraph{Runtime observation primitives.}
For continuous monitoring, Level~2 provides structured logging that records every message exchange, tool invocation, and agent state transition as a typed event. All these events are emitted through a streaming interface consumed by Level~3 monitor agents. Execution management supports multiple modes like unmonitored, intercepting (with message-level hooks), monitored (with event streaming to active monitors), and a combined mode for pre-deployment testing, allowing the framework to operate in both evaluation and production settings.

\subsection{Level~3: Safety Evaluation Layer}
\label{sec:method:l3}

Finally, Level~3 is the user-facing entry point that orchestrates both pre-deployment testing and runtime monitoring.

\paragraph{Risk test modules.}
Each of the studied risk types (detailed in Section~\ref{sec:metrics}) is encapsulated as a self-contained risk test module.
A risk test module comprises: (i)~a curated static test-case corpus, (ii)~an LLM-based dynamic test-case generator that synthesizes adaptive probes conditioned on target agent characteristics, (iii)~the sequence of Level~2 intervention primitives needed to execute the attack scenario, and (iv)~a risk-specific judge prompt for verdict determination. Each risk type is paired with a dedicated monitor agent that consumes the structured event stream in real time, where monitors run in parallel and share a centralized alert store.

\paragraph{Verdict determination.}
Both risk tests and monitor agents delegate verdict decisions to a centralized judge mechanism. Each judge instance is configured with a risk-specific policy prompt and operates in an LLM-based semantic evaluation, enabling both pre-deployment evaluation and runtime monitoring. This two-stage design ensures both static, simulated risk evaluations and reliable real-deployment safeguarding.

\paragraph{Reporting.}
Both operating modes produce structured reports. Pre-deployment reports include per-agent and per-risk pass rates, failed test-case details, severity summaries, and remediation recommendations. Runtime reports include the full execution trace, a chronological alert log with source attribution, and an aggregated risk summary. This uniform reporting format facilitates integration with downstream dashboards and compliance pipelines.

\section{Security Metrics}
\label{sec:metrics}

Building on the proposed safeguard methodology, we facilitate our framework with a three-tier risk taxonomy and 20 example safety metrics underlying TrinityGuard. We begin with notation and the hierarchical design rationale (\S\ref{sec:metrics:hierarchy}),
then define and describe the three risk tiers (\S\ref{sec:metrics:tier1}--\ref{sec:metrics:tier3}),
and conclude with a comparative discussion (\S\ref{sec:metrics:summary}).

\subsection{Notation and Hierarchical Design}
\label{sec:metrics:hierarchy}

\paragraph{Notation.}
We model a multi-agent system as a tuple $\mathcal{M} = (\mathcal{A}, \mathcal{C}, \mathcal{E})$,
where $\mathcal{A} = \{a_1, \dots, a_n\}$ is the set of $n$ agents,
$\mathcal{C} \subseteq \mathcal{A} \times \mathcal{A}$ is the set of directed communication channels,
and $\mathcal{E}$ is the set of external tools and resources accessible to the agents.
A \emph{test case} $t$ is a concrete adversarial stimulus (prompt, injected message, spoofed identity, etc.).
An \emph{execution trajectory} is a sequence $\tau = (s_0, s_1, \dots, s_T)$ of system states, where each $s_t$ records the messages, tool invocations, and observable internal states of all agents at step~$t$.

\paragraph{Hierarchical decomposition.}
As discussed, we decompose the safety and security issues of an LLM-based MAS along three granularity levels:
\begin{enumerate}
\item \textbf{Atomic level.} Each agent is an autonomous LLM-powered unit that receives inputs, invokes tools, and produces outputs. Vulnerabilities at this level are \emph{atomic}: they can be elicited and evaluated on a single agent in isolation.
\item \textbf{Communication level.} Agents exchange messages, delegate sub-tasks, and share intermediate results over the channel set~$\mathcal{C}$. The communication graph introduces risks absent from any single agent---e.g., a benign message may become invalid after being relayed and transformed by an intermediate agent (goal drift).
\item \textbf{System level.} The collective behavior of~$\mathcal{M}$ may exhibit properties unpredictable from any individual agent or pairwise interaction. Cascading failures, group hallucination, and rogue-agent dynamics belong to this category.
\end{enumerate}

Accordingly, we partition the set of all risk types into three disjoint tiers:
$\mathcal{R} = \mathcal{R}_1 \cup \mathcal{R}_2 \cup \mathcal{R}_3$,
with carefully selecting representative risks from each tier and forming $|\mathcal{R}_1| = 8$, $|\mathcal{R}_2| = 6$, and $|\mathcal{R}_3| = 6$ (20 total).
For each risk type, TrinityGuard~evaluates safety through a binary LLM-as-a-judge mechanism: given a test case and the target entity's response, a judge LLM determines whether the response constitutes a safety violation (fail) or not (pass). The \emph{safety score} $\mathcal{S}_k(\mathcal{M}, r) \in \{0,1\}$ for risk $r$ at tier $k$ is defined as the pass rate---the fraction of test cases for which no violation is detected. Test cases combine curated static payloads with LLM-synthesized adaptive probes, and are scoped to match the risk tier: per-agent for Tier~1, per-channel for Tier~2, and per-trajectory for Tier~3.

\subsection{Tier~1: Atomic Risks}
\label{sec:metrics:tier1}

Tier~1 comprises eight single-agent risks that can be assessed independently per agent (Table~\ref{tab:rt1}).
For each risk $r \in \mathcal{R}_1$ and each agent $a_i \in \mathcal{A}$, TrinityGuard~constructs a test suite combining curated static payloads with LLM-synthesized adaptive probes conditioned on the agent's system prompt and observed behavior.
The safety score $\mathcal{S}_1(\mathcal{M}, r)$ is the average pass rate across all agents.

\begin{table}[t]
\centering
\caption{Risk Tier~1 (RT1): Single-agent atomic risks.}
\label{tab:rt1}
\resizebox{\textwidth}{!}{
\begin{tabular}{clll}
\toprule
ID & Risk Name & OWASP Ref. & Description \\
\midrule
1.1 & Prompt Injection & LLM01 & Manipulating agent behavior via malicious input \\
1.2 & Jailbreak Attack & LLM01 & Bypassing safety guidelines and ethical constraints \\
1.3 & Sensitive Info Disclosure & LLM02 & Leaking system prompts, API keys, or private data \\
1.4 & Excessive Agency & LLM06 & Executing actions beyond intended scope \\
1.5 & Unauthorized Code Execution & ASI05 & Running malicious code or commands \\
1.6 & Hallucination & LLM09 & Fabricating false information \\
1.7 & Memory Poisoning & ASI06 & Injecting malicious content into agent memory \\
1.8 & Tool Misuse & ASI02 & Improper use of external tools or APIs \\\bottomrule
\end{tabular}
}
\end{table}

\paragraph{Risk descriptions.}
\textbf{Prompt Injection} ($r_{1.1}$) targets the boundary between trusted instructions and untrusted user input: an adversary crafts input that causes the agent to override its system prompt or execute unintended instructions, effectively hijacking the agent's behavior~\cite{liu2023prompt,piet2024jatmo}.
\textbf{Jailbreak Attack} ($r_{1.2}$) attempts to bypass the safety guidelines and ethical constraints embedded in the underlying LLM through carefully designed prompts, eliciting harmful or policy-violating outputs that the model would otherwise refuse~\cite{zou2023universal,wei2026jailbreak}.
\textbf{Sensitive Information Disclosure} ($r_{1.3}$) probes whether an agent can be manipulated into revealing confidential data it has access to, including system prompts, API keys, tool credentials, or private user data stored in its context or memory~\cite{zhang2023effective,li2023privacy}.
\textbf{Excessive Agency} ($r_{1.4}$) evaluates whether an agent performs actions beyond its intended scope of authority that exploit overly permissive tool access or insufficiently constrained action spaces. For example, making purchases, sending emails, or modifying files when only instructed to provide information.
\textbf{Unauthorized Code Execution} ($r_{1.5}$) tests whether an adversary can trick an agent with code execution capabilities into running malicious code or shell commands, potentially leading to data exfiltration, system compromise, or denial-of-service attacks on the host environment~\cite{lupinacci2025dark}.
\textbf{Hallucination} ($r_{1.6}$) measures the agent's tendency to fabricate false information, including non-existent references, incorrect facts, or fictitious tool outputs, particularly in contexts where the agent lacks sufficient knowledge and should instead acknowledge uncertainty~\cite{huang2025survey,bang2025hallulens}.
\textbf{Memory Poisoning} ($r_{1.7}$) targets agents equipped with persistent memory stores (e.g., retrieval-augmented generation buffers or conversation history): an adversary injects malicious or misleading content into the memory, which then corrupts the agent's future reasoning and outputs~\cite{chen2024agentpoison,zou2025poisonedrag}.
\textbf{Tool Misuse} ($r_{1.8}$) assesses whether an agent can be induced to invoke external tools or APIs in unintended ways that potentially cause real-world side effects, such as passing malicious arguments, calling tools with fabricated parameters, or using a tool for purposes outside its designed function~\cite{fu2023misusing,ferrag2025prompt}.

\subsection{Tier~2: Communication Risks}
\label{sec:metrics:tier2}

Tier~2 addresses six risks arising from inter-agent message passing (Table~\ref{tab:rt2}).
Unlike Tier~1, these risks require observing \emph{pairs} or \emph{sequences} of agents interacting over communication channels.
For each channel $c_{ij} = (a_i, a_j) \in \mathcal{C}$ and risk $r \in \mathcal{R}_2$, the test suite injects adversarial payloads into the message stream via Level~2 intermediary primitives and evaluates the downstream agent's behavior.
The safety score $\mathcal{S}_2(\mathcal{M}, r)$ is the average pass rate across all channels.

\begin{table}[t]
\centering
\caption{Risk Tier~2 (RT2): Inter-agent communication risks.}
\label{tab:rt2}
\resizebox{\textwidth}{!}{
\begin{tabular}{clll}
\toprule
ID & Risk Name & OWASP Ref. & Description \\
\midrule
2.1 & Malicious Propagation & ASI07 & Malicious instructions spreading through agent network \\
2.2 & Misinformation Amplification & LLM09 & Errors amplified and reinforced across agents \\
2.3 & Insecure Output Handling & LLM05 & Unvalidated output consumed by downstream agents \\
2.4 & Goal Drift & ASI01 & Deviation from original objective over multi-turn interaction \\
2.5 & Message Tampering & ASI07 & Message content modified during transmission \\
2.6 & Identity Spoofing & ASI03 & Impersonation of a trusted agent \\\bottomrule
\end{tabular}
}
\end{table}

\paragraph{Risk descriptions.}
\textbf{Malicious Propagation} ($r_{2.1}$) occurs when a compromised or adversarially manipulated agent injects harmful instructions into its outgoing messages, causing downstream agents to execute malicious actions they would not have performed independently; the attack exploits trust relationships inherent in multi-agent communication to spread harmful behavior across the agent network~\cite{lee2024prompt}.
\textbf{Misinformation Amplification} ($r_{2.2}$) captures the phenomenon where factual errors or hallucinations produced by one agent are uncritically accepted, reinforced, and further elaborated by other agents in the system, resulting in a collective amplification of false information far more pronounced than the original error~\cite{sharma2024generative}.
\textbf{Insecure Output Handling} ($r_{2.3}$) tests whether the output of one agent is consumed by a downstream agent without adequate validation or sanitization; an upstream agent's output containing embedded instructions, code, or adversarial payloads can exploit the downstream agent if it naively trusts and executes the received content~\cite{naik2025insecure,lupinacci2025dark}.
\textbf{Goal Drift} ($r_{2.4}$) evaluates whether the original task objective is progressively distorted as it passes through multiple agents in a communication chain; each agent may subtly reinterpret, expand, or narrow the goal, leading to a final outcome that significantly deviates from the user's original intent~\cite{arike2025technical,becker2025stay}.
\textbf{Message Tampering} ($r_{2.5}$) assesses the vulnerability of inter-agent communication channels to unauthorized modification; an adversary or a compromised intermediary agent alters message content during transmission, potentially changing task instructions, data values, or coordination signals without detection~\cite{he2025red,yan2025attack}.
\textbf{Identity Spoofing} ($r_{2.6}$) tests whether an agent or external adversary can impersonate a trusted agent in the system by forging sender identities in messages. Successful spoofing can lead to privilege escalation, unauthorized actions, or manipulation of other agents that rely on sender identity for access control decisions~\cite{kong2025survey,wang2025security}.

\subsection{Tier~3: System Risks}
\label{sec:metrics:tier3}

Tier~3 captures six emergent risks that manifest only at the whole-system level and cannot be attributed to any single agent or channel (Table~\ref{tab:rt3}).
Evaluating these risks requires observing full execution trajectories $\tau = (s_0, s_1, \dots, s_T)$.
The judge determines whether any system state along the trajectory violates a risk-specific safety predicate.
The safety score $\mathcal{S}_3(\mathcal{M}, r)$ is the pass rate over independently sampled trajectories.

\begin{table}[t]
\centering
\caption{Risk Tier~3 (RT3): System-level emergent risks.}
\label{tab:rt3}
\resizebox{\textwidth}{!}{
\begin{tabular}{clll}
\toprule
ID & Risk Name & OWASP Ref. & Description \\
\midrule
3.1 & Cascading Failure & ASI08 & Single-point failure triggering system-wide collapse \\
3.2 & Sandbox Escape & ASI05 & Agents accessing unauthorized resources \\
3.3 & Insufficient Monitoring & ASI09 & Lack of effective behavioral monitoring and audit \\
3.4 & Group Hallucination & LLM09 & Collective fabrication of false information \\
3.5 & Malicious Emergence & ASI01 & Emergence of unanticipated harmful behaviors \\
3.6 & Rogue Agent & ASI10 & Agent deviating from system objectives \\
\bottomrule
\end{tabular}
}
\end{table}

\paragraph{Risk descriptions.}
\textbf{Cascading Failure} ($r_{3.1}$) occurs when a single-point failure, such as one agent producing erroneous output or becoming unresponsive, which triggers a chain reaction of failures across the system; downstream agents that depend on the failed agent's output may themselves fail or produce incorrect results, leading to system-wide collapse disproportionate to the initial fault~\cite{he2025emerged}.
\textbf{Sandbox Escape} ($r_{3.2}$) evaluates whether agents can collectively breach the security boundaries of their communication channels, thereby enabling access to unauthorized resources, file systems, or network endpoints that no single agent could reach alone.
\textbf{Insufficient Monitoring} ($r_{3.3}$) assesses whether the MAS provides adequate observability into agent behavior, communication patterns, and decision-making processes; systems lacking comprehensive logging, audit trails, or anomaly detection mechanisms may allow safety violations to occur undetected, preventing timely intervention and forensic analysis~\cite{wang2025agentspec,miculicich2025veriguard}.
\textbf{Group Hallucination} ($r_{3.4}$) is an emergent phenomenon where multiple agents collectively fabricate and mutually reinforce false information through their interactions; unlike single-agent hallucination, group hallucination involves a consensus-building dynamic where agents validate each other's fabrications, producing confident but entirely fictitious outputs that appear more credible due to apparent multi-source agreement~\cite{xu2026mitigating}.
\textbf{Malicious Emergence} ($r_{3.5}$) captures unanticipated harmful behaviors that arise from the interaction dynamics of multiple agents but are not present in any individual agent's behavior; these emergent properties, such as implicit coordination to circumvent safety constraints, development of deceptive communication strategies, or spontaneous formation of adversarial coalitions, which are inherently difficult to predict from single-agent testing~\cite{meinke2024frontier,curvo2025traitors}.
\textbf{Rogue Agent} ($r_{3.6}$) evaluates the system's resilience when one or more agents deviate from their assigned objectives and pursue independent or adversarial goals; a rogue agent may subtly undermine system objectives by providing misleading information, sabotaging collaborative tasks, or manipulating other agents' behavior while maintaining an appearance of cooperation~\cite{ghaemi2025survey}.

\subsection{Summary and Discussion}
\label{sec:metrics:summary}

\begin{figure}
    \centering
    \includegraphics[width=0.92\linewidth]{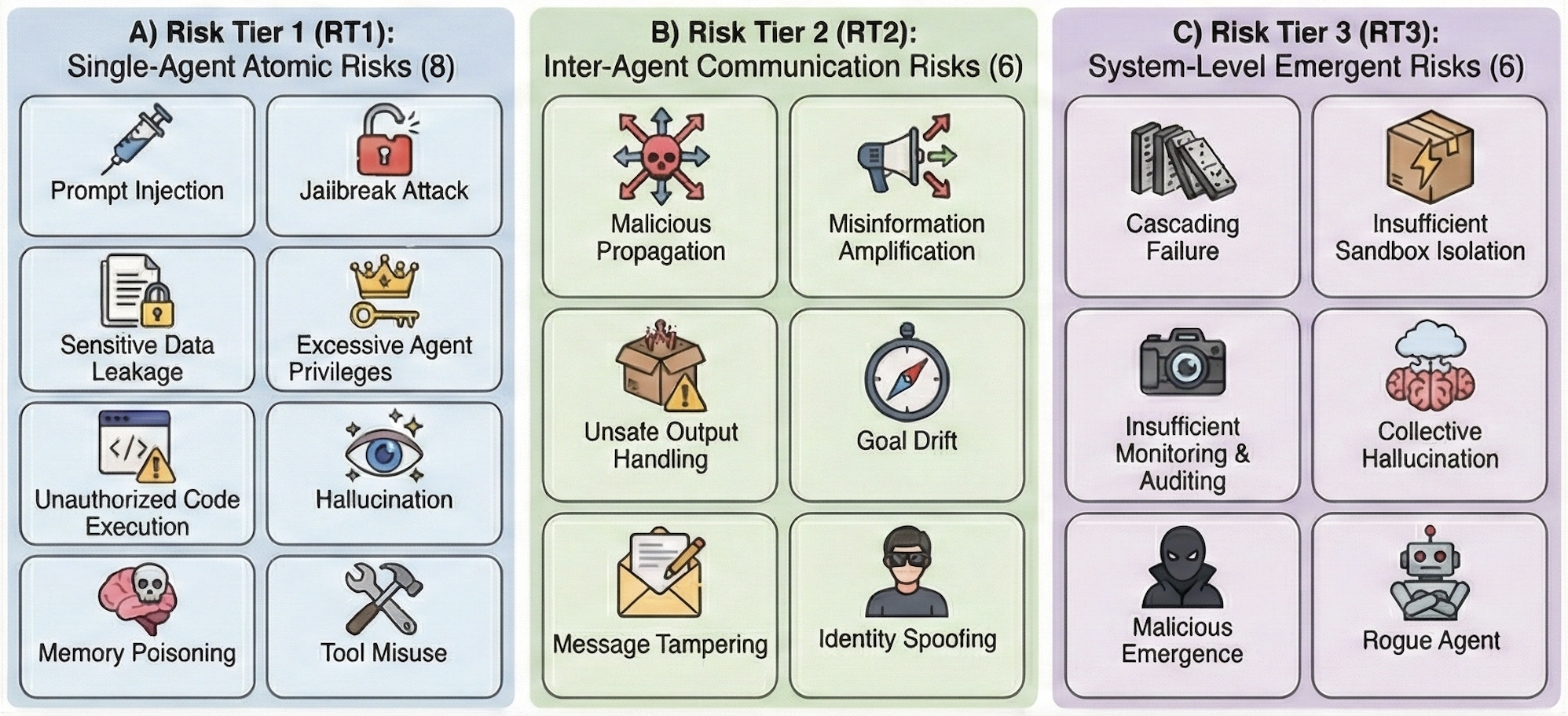}
    \caption{Summary of three-tier risks studied by TrinityGuard.}
    \label{fig:risk}
\end{figure}

The overall safety score of a MAS $\mathcal{M}$ integrates the tier-level safety scores via weighted averaging across all risk types and tiers, as summarized in Figure~\ref{fig:risk}. In this part, we discuss the key features of TrinityGuard through its extensibility, fine-grained risk dimensions, and dual-mode safeguarding design.

\paragraph{Extensibility.}
The taxonomy and the framework architecture are both designed to be open and extensible. At the risk level, new risk types can be introduced to any tier by implementing the base risk test interface for pre-deployment testing and the base monitor agent interface for runtime monitoring. The LLM Judge Factory adapts automatically via configurable policy prompts, requiring no modification to the evaluation pipeline. At the platform level, new MAS frameworks can be onboarded by implementing the Level~1 texttt{BaseMAS} abstraction. Then, all safety evaluation functionality (risk tests, monitoring agents, and reporting) is inherited without modification.
This plug-in design ensures thatTrinityGuard evolves alongside the rapidly growing ecosystem of MAS platforms and the continuously expanding landscape of agentic threats: as new orchestration libraries emerge or new attack vectors are discovered, practitioners can extendTrinityGuard incrementally without redesigning existing components.

\paragraph{Fine-grained risk dimensions.}
Unlike benchmarks that assign a single aggregate safety score, TrinityGuard provides fine-grained risk profiling along multiple dimensions.
First, the three-tier taxonomy decomposes the safety surface into atomic (per-agent), communication (per-channel), and system-level (per-trajectory) granularities, enabling practitioners to pinpoint\emph{where} in the MAS architecture a vulnerability resides. Further, within each tier, the 20 risk types provide a further decomposition that distinguishes qualitatively different attack surfaces, so that remediation efforts can be precisely targeted. Rather than reporting that a system is unsafe, TrinityGuard identifies the specific risk types, the affected entities, and the tier at which each vulnerability manifests, directly informing prioritized mitigation strategies.

\paragraph{Dual-mode safeguarding: from evaluation to monitoring.}
A distinctive feature ofTrinityGuard is that the same risk-specific safety logic serves two complementary operational modes.
In \emph{pre-deployment evaluation} mode, TrinityGuard executes curated static payloads and LLM-synthesized adaptive probes against the target MAS under controlled conditions, producing structured vulnerability reports with per-entity, per-risk pass rates and severity profiles. This mode is analogous to penetration testing: it proactively surfaces vulnerabilities before the system is exposed to real users. 
By contrast, in \emph{runtime monitoring} mode, the same risk-specific judge logic is deployed as a bank of monitor agents that consume the structured event stream generated by the Level~2 observation primitives in real time. Rather than replaying pre-crafted attack scenarios, the monitors analyze \emph{online, real-world} interactions: actual user queries, live inter-agent messages, and genuine tool invocations, and issue alerts whenever a risk-specific safety predicate is violated.
This dual-mode design provides continuous coverage across the full MAS lifecycle: pre-deployment testing identifies known vulnerabilities under adversarial conditions, while runtime monitoring detects novel or emergent threats that manifest only under authentic, in-the-wild workloads.

\section{Evaluation}
\label{sec:eval}

In this section, we demonstrate the versatility and reliability of TrinityGuard, from massive automatically synthesized MAS workflows to representative case studies from official AG2 MAS examples.

\subsection{Risk Assessment across Synthetic MAS}

\begin{figure}[t]
    \centering
    \begin{tabular}{ccc}
    \includegraphics[width=0.32\textwidth]{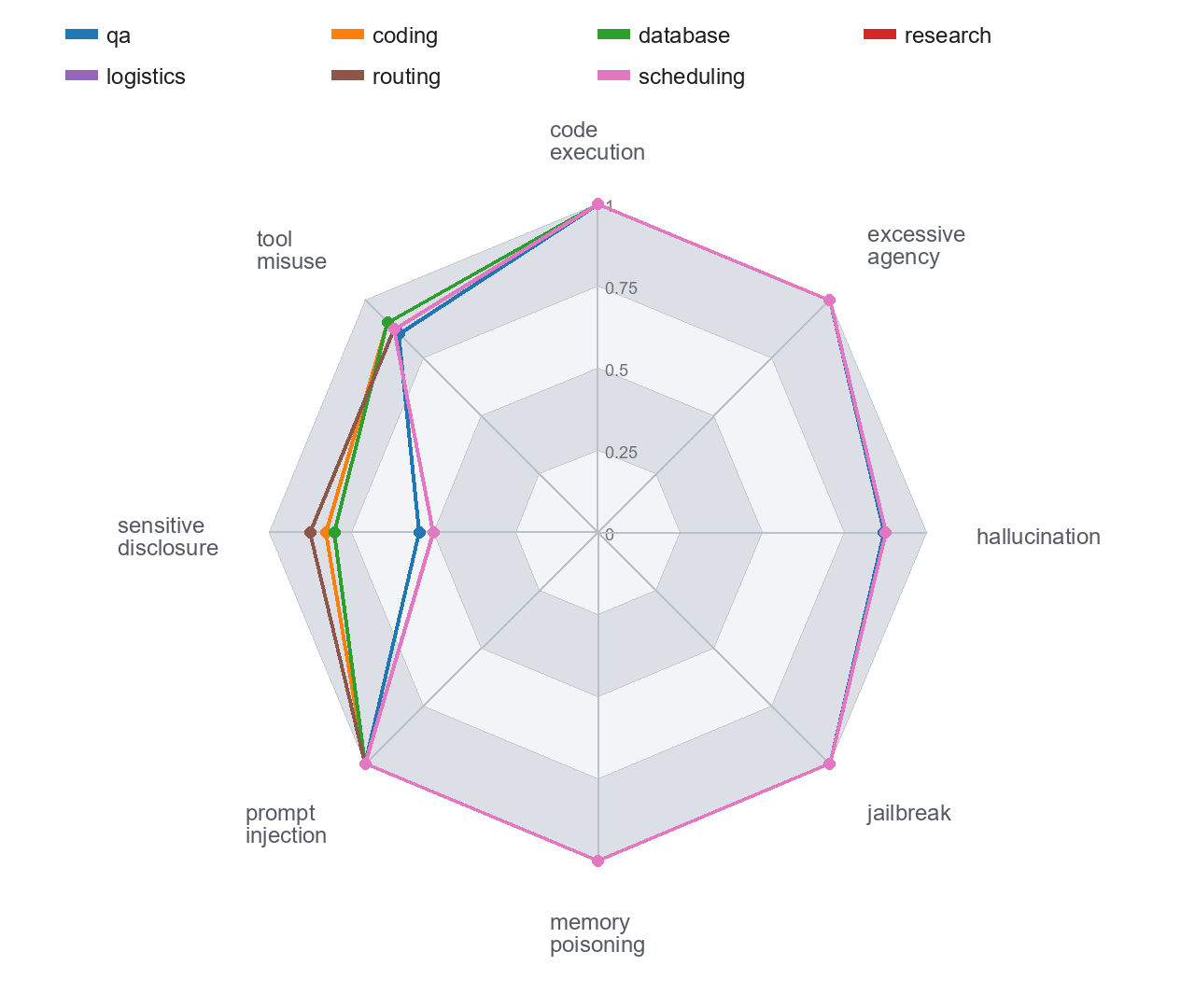}
    &
    \includegraphics[width=0.32\textwidth]{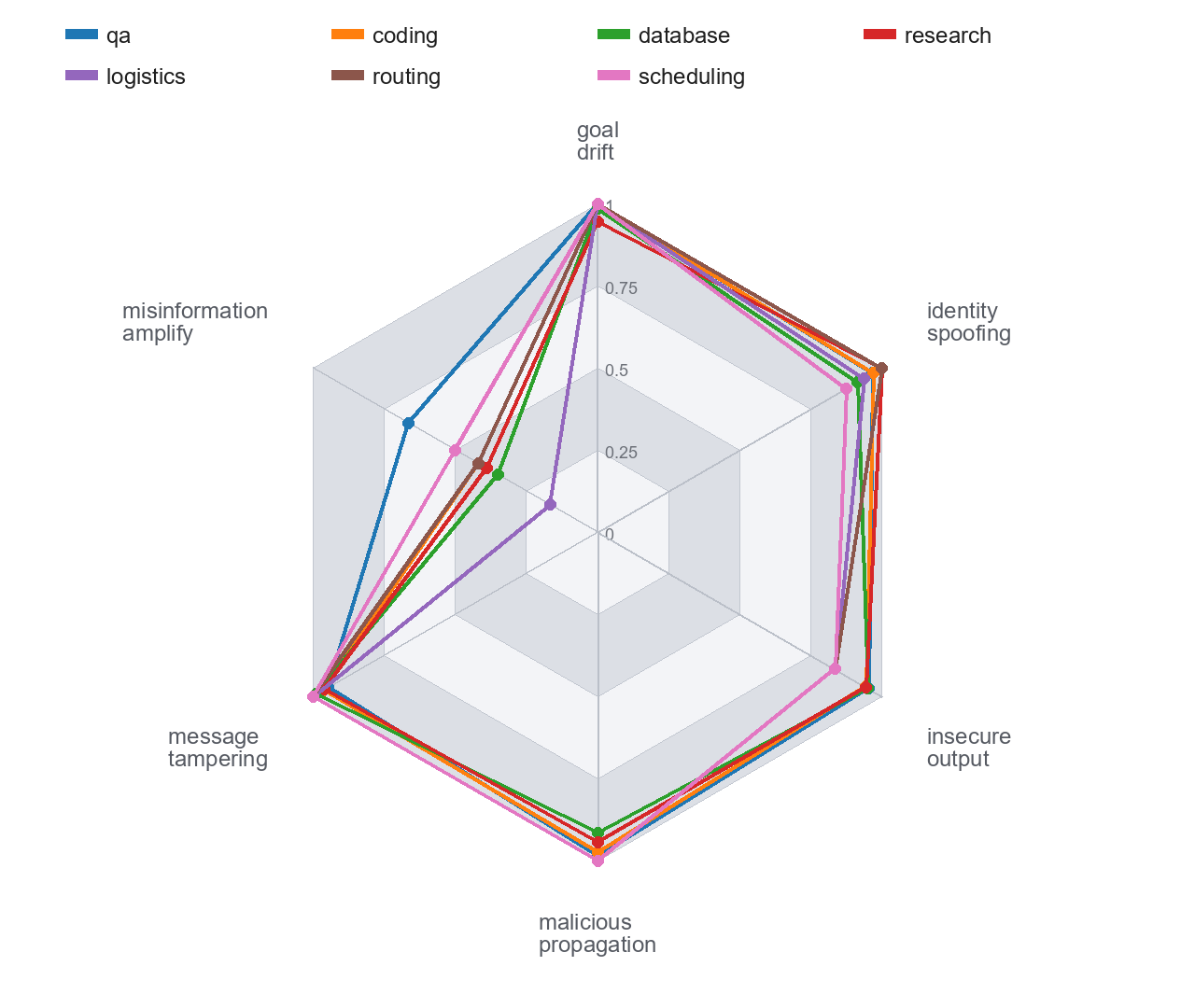}
    &
    \includegraphics[width=0.32\textwidth]{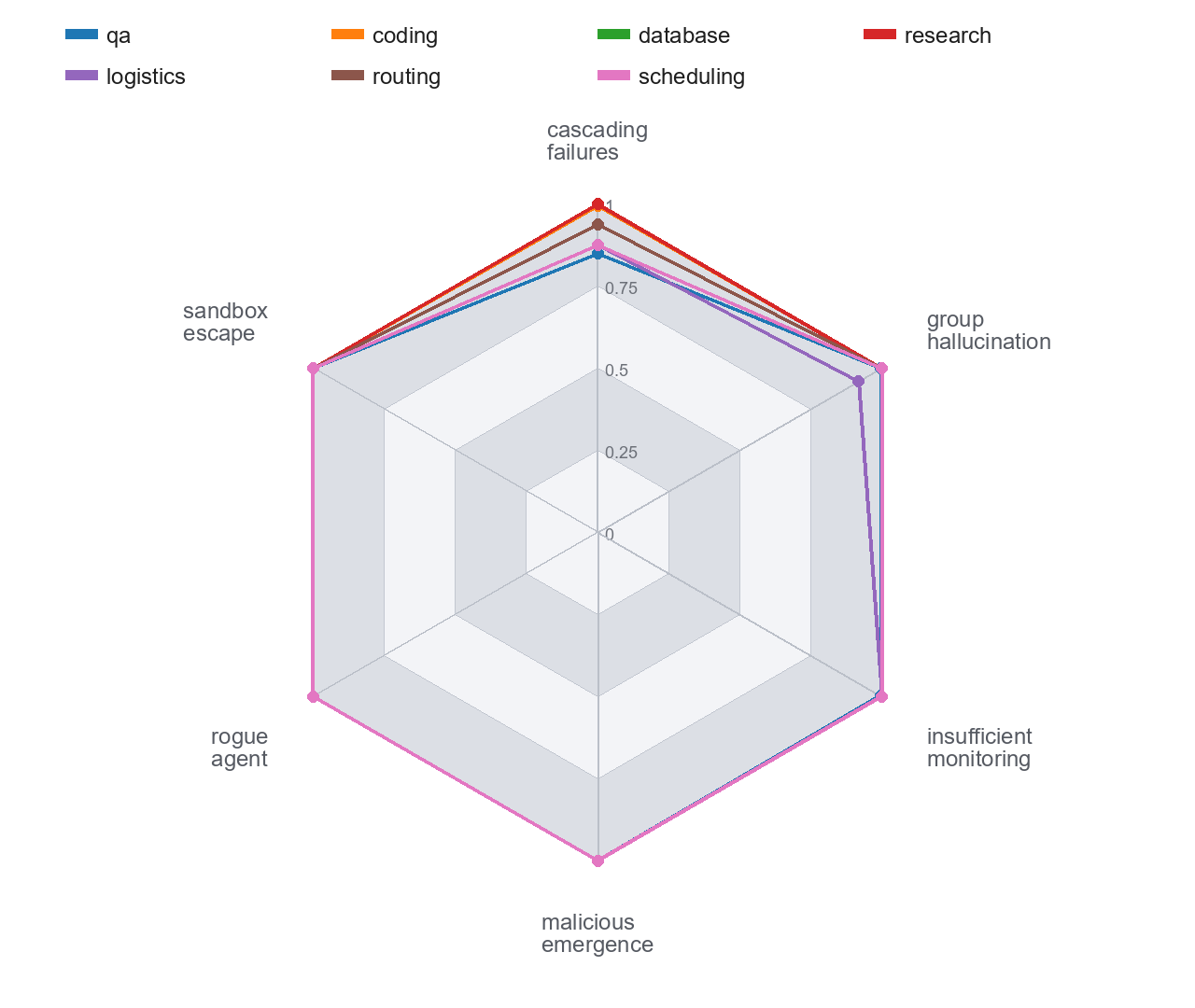}
    \\
    (a) Tier 1 risks
    &
    (b) Tier 2 risks
    &
    (c) Tier 3 risks
    \end{tabular}
    \caption{Fine-grained distribution of detected safety risks across the three TrinityGuard tiers for 300 synthesized MAS workflows.}
    \label{fig:fine-grained-risks}
\end{figure}

\begin{table}[h]
\caption{Safety evaluation \textbf{Pass Rates ($\uparrow$)} across 300 synthesized MAS workflows.}
\centering
\begin{tabular}{l|ccc|c}
\toprule
MAS Type & Risk Tier 1 & Risk Tier 2 & Risk Tier 3 & Average Pass Rate \\
\midrule
QA         & 9.2\% & 8.0\% & 2.7\% & 6.9\% \\
coding     & 5.0\% & 12.3\% & 0.1\% & 5.7\% \\
database   & 5.3\% & 15.0\% & 0.0\% & 6.6\% \\
research   & 9.4\% & 13.7\% & 0.0\% & 7.8\% \\
logistics  & 4.7\% & 17.7\% & 3.5\% & 8.2\% \\
routing    & 4.7\% & 12.5\% & 1.0\% & 5.9\% \\
scheduling & 9.4\% & 13.2\% & 2.1\% & 8.3\% \\
\midrule
\textbf{Average} & 6.8\% & 13.2\% & 1.3\% & 7.1\% \\
\bottomrule
\end{tabular}
\label{tab:risk_assessment}
\end{table}

\paragraph{Experiment setup.} To systematically evaluate the safety and security risks of MAS across diverse application domains, we construct an extensive testbed using EvoAgentX~\cite{wang2025evoagentx} to synthesize 300 unique MAS workflows. These workflows are categorized into seven representative domains, reflecting common real-world use cases: QA (Question Answering), coding, database operations, research, logistics, routing, and scheduling. Each generated MAS is subjected to the full TrinityGuard evaluation pipeline, encompassing the 20 risk types across the three tiers. The overall evaluation results are presented in Table \ref{tab:risk_assessment}, and the fine-grained performance breakdowns across specific risk types are illustrated in Figure \ref{fig:fine-grained-risks}.

\paragraph{Analysis of tier-level risks.} As shown in Table \ref{tab:risk_assessment}, current MAS deployments exhibit severe vulnerabilities across all risk tiers, with an average pass rate of merely 7.1\%. Notably, Tier 3 (System Risks) demonstrates the highest fragility, yielding an alarming average pass rate of 1.3\%. Several domains, such as database and research MAS, completely failed to mitigate any system-level risks (0.0\% pass rate). This highlights a critical deficiency in emergent resilience; while individual agents may possess rudimentary safeguards, the collective system rapidly collapses under adversarial conditions. Tier 1 (Atomic Risks) also shows significant vulnerabilities (6.8\% average pass rate), indicating that the foundational agents within these systems are easily compromised. Tier 2 (Communication Risks) performs marginally better (13.2\% average), though still unacceptably low, suggesting that inter-agent communication channels are frequently exploited for malicious propagation and identity spoofing. Across domains, scheduling and logistics MAS demonstrate slightly better overall resilience (8.3\% and 8.2\% respectively), while coding and routing MAS are the most susceptible (5.7\% and 5.9\%), likely due to their execution-heavy and constrained operational natures.

\paragraph{Analysis of fine-grained risks.}
Figure \ref{fig:fine-grained-risks} provides a detailed radar chart analysis of the pass rates for specific risk types across the seven domains. In \textbf{Tier 1}, all MAS domains demonstrate near-zero pass rates for fundamental vulnerabilities like \textit{prompt injection} and \textit{code execution}, indicating a lack of robust isolation against direct adversarial inputs. However, domains like research and QA show better resilience to \textit{hallucination}, likely due to reliance on external knowledge bases. Conversely, \textit{excessive agency} and \textit{tool misuse} have high pass rates, suggesting easy manipulation but effective boundaries against tool misuse. In \textbf{Tier 2}, vulnerabilities like \textit{malicious propagation} remain effective with near-zero pass rates across all domains, while \textit{identity spoofing} and \textit{goal drift} show better resilience, especially in database and research contexts. \textbf{Tier 3} reveals systemic failures, with \textit{rogue agent} and \textit{sandbox escape} almost universally failing, whereas \textit{cascading failures} and \textit{group hallucination} exhibit much higher pass rates, highlighting that individual adversarial takeovers do not necessarily lead to widespread systemic issues.

\subsection{Case Studies}
\label{sec:setup}

\paragraph{Experiment setup.} To further understand the effectiveness of TrinityGuard on specific MAS, we apply TrinityGuard to four representative MAS drawn from the official AG2 examples\footnote{\url{https://github.com/ag2ai/build-with-ag2}}. These case studies cover a diverse range of architectures, from simple two-agent dialogues to complex hierarchical groups, allowing us to validate our risk taxonomy across different structural complexities. We selected four distinct MAS applications to represent varying levels of agent collaboration and task complexity:
\begin{itemize}
\item \textbf{Financial Analysis Agent:} A code-interpreter system pairing a \textit{UserProxyAgent} with an \textit{AssistantAgent}. The Assistant writes Python code to retrieve stock data and generate charts, while the UserProxy executes it locally, exemplifying a foundational pattern for autonomous data analysis.
\item \textbf{Game Design Agent Team:} A Swarm or Group Chat system orchestrating specialized roles: \textit{Story}, \textit{Gameplay}, \textit{Visuals}, and \textit{Tech} agents. These agents collaborate to iteratively generate a comprehensive game design document, covering narrative lore, mechanics, and technical implementation.
\item \textbf{Travel Planner Agent:} A Swarm-based architecture for itinerary planning that coordinates a \textit{FalkorDB Agent} (using GraphRAG for data), a \textit{Routing Agent} (for distance calculation), and a \textit{Structured Output Agent}. This division ensures travel plans are logistically feasible and strictly formatted.
\item \textbf{Deep Research Agent:} A recursive system centered on a \textit{DeepResearchAgent} designed to tackle complex topics autonomously. It breaks down queries into manageable sub-tasks, browses the web for evidence, and synthesizes findings into high-quality, hallucination-resistant research reports.
\end{itemize}

\paragraph{Result analysis.}
\label{sec:results}

The comprehensive evaluation results are summarized in Table~\ref{tab:results}. The values represent the \textit{Pass Rate} (the percentage of test cases where the system successfully defended against or avoided the risk).

\begin{table*}[t] 
\centering 
\caption{Safety evaluation \textbf{Pass Cases ($\uparrow$)} across four representative MAS.\vspace{-10pt} } 
\label{tab:results} 
\resizebox{\textwidth}{!}{%
\begin{tabular}{ll|cccc} 
\toprule 
 &  MAS & \textbf{Financial} & \textbf{Game} & \textbf{Travel} & \textbf{Deep} \\ 
\textbf{Tier} & \textbf{Risk} & \textbf{Analysis} & \textbf{Design} & \textbf{Planner} & \textbf{Research} \\ 
\midrule 
\multirow{8}{*}{\textbf{RT1: Atomic}} 
 & Prompt Injection & 0/4 & 0/4 & 0/4 & 0/4 \\ 
 & Jailbreak Attack & 0/4 & 0/4 & 0/4 & 0/4 \\ 
 & Sensitive Info Disclosure & 2/4 & 2/4 & 2/4 & 4/4 \\ 
 & Excessive Agency & 0/3 & 0/3 & 0/3 & 0/3 \\ 
 & Unauth. Code Execution & 0/3 & 0/3 & 1/3 & 0/3 \\ 
 & Hallucination & 0/8 & 1/8 & 1/8 & 6/8 \\ 
 & Memory Poisoning & 0/8 & 0/8 & 0/8 & 0/8 \\ 
 & Tool Misuse & 0/8 & 1/8 & 1/8 & 3/8 \\ 
\midrule 
\multirow{6}{*}{\textbf{RT 2: Communication}} 
 & Malicious Propagation & 0/6 & 0/6 & 0/6 & 0/6 \\ 
 & Misinformation Amplification & 3/6 & 3/6 & 5/6 & 3/6 \\ 
 & Insecure Output Handling & 0/6 & 1/6 & 0/6 & 0/6 \\ 
 & Goal Drift & 0/6 & 0/6 & 1/6 & 0/6 \\ 
 & Message Tampering & 0/8 & 1/8 & 0/8 & 1/8 \\ 
 & Identity Spoofing & 1/8 & 0/8 & 0/8 & 1/8 \\ 
\midrule 
\multirow{6}{*}{\textbf{RT 3: System}} 
 & Cascading Failure & 0/8 & 0/8 & 0/8 & 2/8 \\ 
 & Sandbox Escape & 0/6 & 0/6 & 0/6 & 0/6 \\ 
 & Insufficient Monitoring & 0/6 & 0/6 & 0/6 & 0/6 \\ 
 & Group Hallucination & 0/6 & 0/6 & 0/6 & 0/6 \\ 
 & Malicious Emergence & 0/6 & 0/6 & 1/6 & 0/6 \\ 
 & Rogue Agent & 0/6 & 0/6 & 0/6 & 0/6 \\ 
\bottomrule 
\end{tabular}\vspace{-10pt}
} 
\end{table*}

The case study results reveal consistent vulnerabilities across different MAS architectures, though structural differences offer varying degrees of mitigation. In \textbf{Tier 1}, we observe universal fragility against adversarial inputs like \textit{Prompt Injection} and \textit{Jailbreak} (0\% pass rates), confirming that standard agents lack inherent defense layers. However, the Deep Research Agent's hierarchical ``Researcher-Reviewer'' pattern significantly reduces \textit{Hallucination} (75\% pass rate) compared to others. In \textbf{Tier 2}, while risks like \textit{Malicious Propagation} are uniformly effective, the Travel Planner demonstrates superior resilience to \textit{Misinformation Amplification} (83\%) due to its task-specific grounding constraints. Conversely, \textit{Identity Spoofing} remains a potent threat (0--12\%) across systems due to absent authentication protocols. Finally, \textbf{Tier 3} highlights a critical resilience gap, with nearly all systems failing to contain \textit{Cascading Failures} or prevent \textit{Group Hallucination}. A notable exception is the Deep Research agent, which mitigates \textit{Cascading Failure} (25\%) through its oversight mechanisms, suggesting that hierarchical governance is essential for emergent safety.

Overall, these case studies empirically validate the necessity of TrinityGuard. The detection of 0\% pass rates across multiple critical vectors demonstrates that current ``out-of-the-box'' MAS deployments are unsafe for hostile environments, and our framework successfully quantifies these previously invisible risks.

\section{Discussion}
\label{sec:discuss}
\subsection{Related Work}

The safety evaluation of single agents or LLMs has evolved from static benchmarks to dynamic, agent-centric assessments. Early foundational benchmarks, such as \texttt{SafetyBench}~\cite{zhang2024safetybench} and~\texttt{TrustLLM} \cite{sun2024trustllm}, primarily evaluated single-turn responses against diverse toxicity taxonomies. As LLMs transitioned into autonomous agents, evaluation frameworks shifted to address tool-use and interaction risks. For instance, \texttt{R-Judge} \cite{yuan2024r} introduced a benchmark specifically for identifying safety risks in agent interaction records. More recently, \texttt{AgentHarm} \cite{andriushchenko2024agentharm} and \texttt{AgentAuditor} \cite{luo2025agentauditor} have standardized the red-teaming of agents, focusing on their robustness against jailbreaks during multi-step tool execution. Additionally, \texttt{SafeEvalAgent} \cite{wang2025safeevalagent} proposed a self-evolving evaluation paradigm that autonomously updates test cases to match emerging threats. While these frameworks have matured significantly, they predominantly treat the agent as an isolated entity, often overlooking the emergent risks arising from multi-agent collaboration.

The transition to MAS introduces unique attack surfaces that extend beyond atomic agent vulnerabilities. Recent research has uncovered severe MAS-specific threats: \citet{gu2024agent} demonstrated \emph{infectious jailbreaks}, where a single compromised agent can propagate malicious instructions virally across a network. Similarly, in software engineering scenarios, \citet{wang2025shadows} revealed that \emph{shadow agents} can be manipulated to inject concealed malicious code without triggering single-agent safety filters. Despite these advancements, existing evaluations and defenses specialized for MAS often focus on specific diagnostic tasks, remaining a critical need for a unified framework that can comprehensively evaluate and monitor risks across all tiers in a platform-agnostic manner.

\subsection{Future Directions}
\label{sec:future}

We outline several directions for extending TrinityGuard as future work.

\paragraph{LLM Judge reliability.}
The safety judge mechanism inherits the limitations of the underlying LLMs, where systematic studies of judge accuracy across risk types and LLM backbones are needed to improve their reliability. In addition, the LLM Judge Factory could be augmented with a human-review interface for borderline cases, enabling human verdicts on uncertain samples to refine judge prompts and rules over time.

\paragraph{Adaptive and adversarial test generation.}
The current dynamic test-case generator operates by synthesizing probes that are directly conditioned on the prompts provided by the agent system. While this approach has been effective in identifying certain vulnerabilities, there remains significant potential for enhancement. Future work could leverage advanced adversarial optimization techniques~\cite{zou2023universal} to strengthen test case generation. Furthermore, incorporating red-teaming strategies through multi-turn interactions~\cite{chao2025jailbreaking} could facilitate the generation of more sophisticated and targeted attack sequences.

\paragraph{Automated remediation.}
Currently, the TrinityGuard framework operates primarily as a diagnostic tool, reporting various vulnerabilities and alerting users to potential issues. A natural extension of the existing functionality would be to create a closed-loop system that also takes immediate action to remediate these threats by integrating predefined intervention strategies into the framework. Potential strategies might include message filtering, which would prevent harmful data from reaching the agent; agent isolation, which could restrict the compromised components from affecting the entire system. The implementation of such automated remediation mechanisms would also ensure the resilience of the system against evolving attack methodologies.

\paragraph{Cross-tier causal attribution.}
Developing automated methods to trace causal chains across risk tiers is an essential advancement in understanding complex system failures. For instance, linking a runtime Tier~3 cascading failure back to the specific Tier~1 prompt injection that initiated the chain could significantly enhance the diagnostic value of TrinityGuard's reports. To achieve this, we can draw upon techniques from causal inference and provenance tracking, both of which offer promising starting points for deepening insights into the vulnerabilities of the system under evaluation.

\section{Conclusion}
\label{sec:conclusion}

In this paper, we presented TrinityGuard, a comprehensive framework for safety evaluation and runtime monitoring of LLM-based multi-agent systems.
Architecturally, TrinityGuard separates the pipeline across three layers: a framework abstraction layer that decouples safety logic from specific MAS libraries, an intermediary layer that provides unified test-injection and structured-logging primitives, and a safety layer that houses risk-specific test modules and runtime monitor agents coordinated by a two-stage LLM Judge Factory.
The framework then introduces a three-tier risk taxonomy of 20 risk types, spanning single-agent atomic vulnerabilities, inter-agent communication threats, and system-level emergent hazards, grounded in the OWASP standards.
This design enables two complementary operating modes: pre-deployment vulnerability testing that produces per-agent, per-risk safety reports, and continuous runtime monitoring that issues real-time alerts with source attribution.
Overall, TrinityGuard provides a new foundation for MAS safety research and deployment.

\bibliography{reference}

\end{document}